

\def\singlespace{\normalbaselines}
\def\oneandahalfspace{\baselineskip=1.15\normalbaselineskip plus 1pt
\lineskip=2pt\lineskiplimit=1pt}

\def\np{\vfill\eject}
\def\nl{\hfil\break}

\def\nofirstpagenoten{\nopagenumbers\footline={\ifnum\pageno>1\tenrm
\hss\folio\hss\fi}}
\def\nofirstpagenotwelve{\nopagenumbers\footline={\ifnum\pageno>1\twelverm
\hss\folio\hss\fi}}
\def\leaderfill{\leaders\hbox to 1em{\hss.\hss}\hfill}
\def\ft#1#2{{\textstyle{{#1}\over{#2}}}}
\def\frac#1/#2{\leavevmode\kern.1em
\raise.5ex\hbox{\the\scriptfont0 #1}\kern-.1em/\kern-.15em
\lower.25ex\hbox{\the\scriptfont0 #2}}
\def\sfrac#1/#2{\leavevmode\kern.1em
\raise.5ex\hbox{\the\scriptscriptfont0 #1}\kern-.1em/\kern-.15em
\lower.25ex\hbox{\the\scriptscriptfont0 #2}}


\parindent=20pt
\def\narrow{\advance\leftskip by 40pt \advance\rightskip by 40pt}

\def\AB{\bigskip
        \centerline{\bf ABSTRACT}\medskip\narrow}
\def\nonarrower{\advance\leftskip by -40pt\advance\rightskip by -40pt}
\def\AE{\bigskip\nonarrower}

\def\boxit#1{\vbox{\hrule\hbox{\vrule\kern3pt
        \vbox{\kern3pt#1\kern3pt}\kern3pt\vrule}\hrule}}

\def\gtorder{\mathrel{\raise.3ex\hbox{$>$}\mkern-14mu
             \lower0.6ex\hbox{$\sim$}}}
\def\ltorder{\mathrel{\raise.3ex\hbox{$<$}|mkern-14mu
             \lower0.6ex\hbox{\sim$}}}
\def\dalemb#1#2{{\vbox{\hrule height .#2pt
        \hbox{\vrule width.#2pt height#1pt \kern#1pt
                \vrule width.#2pt}
        \hrule height.#2pt}}}

\font\fourteentt=cmtt10 scaled \magstep2
\font\fourteenbf=cmbx12 scaled \magstep1
\font\fourteenrm=cmr12 scaled \magstep1
\font\fourteeni=cmmi12 scaled \magstep1
\font\fourteenss=cmss12 scaled \magstep1
\font\fourteensy=cmsy10 scaled \magstep2
\font\fourteensl=cmsl12 scaled \magstep1
\font\fourteenex=cmex10 scaled \magstep2
\font\fourteenit=cmti12 scaled \magstep1
\font\twelvett=cmtt10 scaled \magstep1 \font\twelvebf=cmbx12
\font\twelverm=cmr12 \font\twelvei=cmmi12
\font\twelvess=cmss12 \font\twelvesy=cmsy10 scaled \magstep1
\font\twelvesl=cmsl12 \font\twelveex=cmex10 scaled \magstep1
\font\twelveit=cmti12
\font\tenss=cmss10
 
 \font\ninebf=cmbx7 scaled \magstep1
\font\ninerm=cmr7 scaled \magstep1 \font\ninei=cmmi7 scaled \magstep1
\font\ninesy=cmsy7 scaled \magstep1 
\font\eightrm=cmr7 scaled 1140 
 
\font\sevenbf=cmbx7 \font\sevenrm=cmr7 \font\seveni=cmmi7
\font\sevensy=cmsy7 

\catcode`@=11
\newskip\ttglue
\newfam\ssfam

\def\fourteenpoint{\def\rm{\fam0\fourteenrm}
\textfont0=\fourteenrm \scriptfont0=\tenrm \scriptscriptfont0=\sevenrm
\textfont1=\fourteeni \scriptfont1=\teni \scriptscriptfont1=\seveni
\textfont2=\fourteensy \scriptfont2=\tensy \scriptscriptfont2=\sevensy
\textfont3=\fourteenex \scriptfont3=\fourteenex \scriptscriptfont3=\fourteenex
\def\it{\fam\itfam\fourteenit} \textfont\itfam=\fourteenit
\def\sl{\fam\slfam\fourteensl} \textfont\slfam=\fourteensl
\def\bf{\fam\bffam\fourteenbf} \textfont\bffam=\fourteenbf
\scriptfont\bffam=\tenbf \scriptscriptfont\bffam=\sevenbf
\def\tt{\fam\ttfam\fourteentt} \textfont\ttfam=\fourteentt
\def\ss{\fam\ssfam\fourteenss} \textfont\ssfam=\fourteenss
\tt \ttglue=.5em plus .25em minus .15em
\normalbaselineskip=16pt
\abovedisplayskip=16pt plus 4pt minus 12pt
\belowdisplayskip=16pt plus 4pt minus 12pt
\abovedisplayshortskip=0pt plus 4pt
\belowdisplayshortskip=9pt plus 4pt minus 6pt
\parskip=5pt plus 1.5pt
\setbox\strutbox=\hbox{\vrule height12pt depth5pt width0pt}
\let\sc=\tenrm
\let\big=\fourteenbig \normalbaselines\rm}
\def\fourteenbig#1{{\hbox{$\left#1\vbox to12pt{}\right.\n@space$}}}

\def\twelvepoint{\def\rm{\fam0\twelverm}
\textfont0=\twelverm \scriptfont0=\ninerm \scriptscriptfont0=\sevenrm
\textfont1=\twelvei \scriptfont1=\ninei \scriptscriptfont1=\seveni
\textfont2=\twelvesy \scriptfont2=\ninesy \scriptscriptfont2=\sevensy
\textfont3=\twelveex \scriptfont3=\twelveex \scriptscriptfont3=\twelveex
\def\it{\fam\itfam\twelveit} \textfont\itfam=\twelveit
\def\sl{\fam\slfam\twelvesl} \textfont\slfam=\twelvesl
\def\bf{\fam\bffam\twelvebf} \textfont\bffam=\twelvebf
\scriptfont\bffam=\ninebf \scriptscriptfont\bffam=\sevenbf
\def\tt{\fam\ttfam\twelvett} \textfont\ttfam=\twelvett
\def\ss{\fam\ssfam\twelvess} \textfont\ssfam=\twelvess
\tt \ttglue=.5em plus .25em minus .15em
\normalbaselineskip=14pt
\abovedisplayskip=14pt plus 3pt minus 10pt
\belowdisplayskip=14pt plus 3pt minus 10pt
\abovedisplayshortskip=0pt plus 3pt
\belowdisplayshortskip=8pt plus 3pt minus 5pt
\parskip=3pt plus 1.5pt
\setbox\strutbox=\hbox{\vrule height10pt depth4pt width0pt}
\let\sc=\ninerm
\let\big=\twelvebig \normalbaselines\rm}
\def\twelvebig#1{{\hbox{$\left#1\vbox to10pt{}\right.\n@space$}}}

\def\tenpoint{\def\rm{\fam0\tenrm}
\textfont0=\tenrm \scriptfont0=\sevenrm \scriptscriptfont0=\fiverm
\textfont1=\teni \scriptfont1=\seveni \scriptscriptfont1=\fivei
\textfont2=\tensy \scriptfont2=\sevensy \scriptscriptfont2=\fivesy
\textfont3=\tenex \scriptfont3=\tenex \scriptscriptfont3=\tenex
\def\it{\fam\itfam\tenit} \textfont\itfam=\tenit
\def\sl{\fam\slfam\tensl} \textfont\slfam=\tensl
\def\bf{\fam\bffam\tenbf} \textfont\bffam=\tenbf
\scriptfont\bffam=\sevenbf \scriptscriptfont\bffam=\fivebf
\def\tt{\fam\ttfam\tentt} \textfont\ttfam=\tentt
\def\ss{\fam\ssfam\tenss} \textfont\ssfam=\tenss
\tt \ttglue=.5em plus .25em minus .15em
\normalbaselineskip=12pt
\abovedisplayskip=12pt plus 3pt minus 9pt
\belowdisplayskip=12pt plus 3pt minus 9pt
\abovedisplayshortskip=0pt plus 3pt
\belowdisplayshortskip=7pt plus 3pt minus 4pt
\parskip=0.0pt plus 1.0pt
\setbox\strutbox=\hbox{\vrule height8.5pt depth3.5pt width0pt}
\let\sc=\eightrm
\let\big=\tenbig \normalbaselines\rm}
\def\tenbig#1{{\hbox{$\left#1\vbox to8.5pt{}\right.\n@space$}}}
\let\rawfootnote=\footnote \def\footnote#1#2{{\rm\parskip=0pt\rawfootnote{#1}
{#2\hfill\vrule height 0pt depth 6pt width 0pt}}}

\def\tenfoot{\tenpoint\hskip-\parindent\hskip-.1cm}

\overfullrule=0pt
\twelvepoint
\def\sbullet{\raise.2em\hbox{$\scriptscriptstyle\bullet$}}
\nofirstpagenotwelve
\hsize=16.5 truecm
\baselineskip 15pt

\def\ft#1#2{{\textstyle{{#1}\over{#2}}}}

\def\a{\alpha_0}

\def\del{\partial}

\def\Z{\ss \rlap Z\mkern3mu Z}

\def\phys{\big|\hbox{phys}\big\rangle}
\def\acrit{\alpha_0^*}

\oneandahalfspace
\rightline{CTP TAMU--5/92}
\rightline{Preprint-KUL-TF-92/1}
\rightline{January 1992}

\vskip 2truecm
\centerline{\bf The Complete Spectrum of the $W_N$ String}
\vskip 1.5truecm
\centerline{H. Lu,$^*$ C.N. Pope,\footnote{$^*$}{\tenfoot Supported in part
by the U.S. Department of Energy, under
grant DE-FG05-91ER40633.} S. Schrans\footnote{$^\diamond$}{\tenfoot
Onderzoeker I.I.K.W.;
On leave of absence from the Instituut voor Theoretische Fysica, \nl
\indent$\,$ K.U. Leuven, Belgium.
}
and K.W.
Xu.\footnote{$^\$ $}{\tenfoot Supported by a World Laboratory
Scholarship.}}
\vskip 1.5truecm
\centerline{\it Center
for Theoretical Physics,
Texas A\&M University,}
\centerline{\it College Station, TX 77843--4242, USA.}

\vskip 1.5truecm
\AB\singlespace
      We obtain the complete physical spectrum of the $W_N$ string, for
arbitrary $N$. The $W_N$ constraints freeze $N-2$ coordinates, while the
remaining coordinates appear in the currents only {\it via} their
energy-momentum tensor.  The spectrum is then effectively described by a set
of ordinary Virasoro-like string theories, but with a non-critical value for
the central charge and a discrete set of non-standard values for the spin-2
intercepts.  In particular, the physical spectrum of the $W_N$ string
includes the usual massless states of the Virasoro string.  By looking at
the norms of low-lying states, we find strong indications that all the $W_N$
strings are unitary.
\AE\oneandahalfspace

\vskip 1.5truecm
\centerline{\tenfoot Available from hepth@xxx/9201050}

\np
\noindent
{\bf 1. Introduction}
\bigskip

     String theory is two-dimensional gravity coupled to a critical matter
system that includes free scalar fields which are interpreted as
coordinates on the target spacetime.  The gauge-fixed string action has a
Virasoro symmetry, which provides a powerful organising principle for
describing the physical spectrum and interactions of the theory.  The
Virasoro algebra is the simplest example of a more general class of
infinite-dimensional symmetry algebras, known generically as $W$ algebras,
which can be characterised by the fact that they contain higher-spin
currents as well as the spin-2 energy-momentum tensor $T(z)$ which generates
the Virasoro subalgebra.  One may then construct generalisations of
two-dimensional gravity, by gauging matter systems with $W$ symmetries.  If
the matter systems are critical, and include free scalars, then one obtains
$W$-string theories.  Owing to the non-linearity of $W$ algebras, the
construction of these $W$-string theories seems to be much more complicated
than that of the usual Virasoro string.  Some progress has been made
recently in constructing specific examples, principally for the case of
$W_3$ [1,2,3].

     $W_3$ is the first example in the infinite sequence of non-linear $W_N$
algebras, which are generated by primary currents of spins $3,\ldots,N$,
together with the energy-momentum tensor $T(z)$.  In this paper, we study
the general features of $W_N$ strings for arbitrary $N$.  Realisations of
$W_N$ in terms of $(N-1)$ free scalars $\varphi_2,\ldots,\varphi_N$ can be
obtained from the Miura transformation for $su(N)$ [4].  These realisations
can be generalised by observing that the scalar $\varphi_2$ enters the
currents only {\it via}\ its energy-momentum tensor (with background
charge), which may then be replaced by an arbitrary energy-momentum tensor
with the same central charge [5,2].  If the new energy-momentum tensor is
chosen to comprise $D$ free scalar fields $X^\mu$ together with the original
scalar $\varphi_2$, then these $(D+1)$ scalars, together with
$\varphi_3,\ldots,\varphi_N$, form the coordinates on the target spacetime.
The $W_N$ constraints, however, ``freeze'' the momentum components of
$\varphi_3,\ldots,\varphi_N$ for physical states to certain specific values.
This implies that only the scalars $\varphi_2$ and $X^\mu$ are
physically-observable coordinates, thus describing a $(D+1)$-dimensional
spacetime.  Furthermore, as we shall show, higher-level physical states in
the theory can only involve excitations in the (unfrozen) $(\varphi_2,
X^\mu)$ directions.  This is because states with excitations in the (frozen)
$(\varphi_3,\ldots,\varphi_N)$ directions turn out to have momentum
components in these directions that are incompatible with momentum
conservation, and thus all such states have zero norm and are to be set
equal to zero.

     Because the physically-observable coordinates $\varphi_2$ and $X^\mu$
enter the $W_N$ currents only through their energy-momentum tensor $T^{\rm
eff}$, the above considerations imply that the $W_N$-string theories
closely resemble ordinary string theory. The central charge $c^{\rm eff}$ of
$T^{\rm eff}$ is related to the total central charge $c_N$ of the $W_N$
realisation.  Requiring that $c_N$ take the critical value
$$
c_N^*=2(N-1)(2N^2+2N+1)\eqno(1.1)
$$
implies that $c^{\rm eff}$ for $W_N$ should take the value
$$
c^{\rm eff}=25+{6\over N(N+1)}.\eqno(1.2)
$$
This suggests that there may be a connection with a Virasoro minimal
model [2], in the sense that $c^{\rm eff}$ may be rewritten as
$$
c^{\rm eff}=26-\Big(1-{6\over N(N+1)}\Big),\eqno(1.3)
$$
where 26 is the critical central charge of the usual Virasoro string and the
term between brackets is precisely the central charge of the unitary
$(N,N+1)$ Virasoro minimal model.

     In this paper, we show how the tachyonic physical states of the $W_N$
string may be classified and generated by the action of the Weyl group of
$su(N)$.  This enables us to explain some findings of [2], where it was
noticed in the special cases of $W_3$ and $W_4$ that certain ``diagonal''
states in the Kac tables of the corresponding minimal models arose at this
tachyonic level.  Our results for the tachyonic states apply to any
$W_N$-string theory, revealing a connection not only with the diagonal
states of the $(N,N+1)$ Virasoro minimal model, but also with the diagonal
states of certain $W_M$ minimal models, for all $M<N$.  We also find a
connection between higher-level states of the $W_N$ string and certain
``off-diagonal'' entries in the Kac table of the relevant minimal model.
However, these particular higher-level states involve excitations in the
frozen directions $(\varphi_3,\ldots,\varphi_N)$, and, as we mentioned
earlier, they therefore have zero norm.

     Because the higher-level states with excitations in these frozen
directions have zero norm, it follows that the only physical states in the
spectrum of the $W_N$ string are the tachyons described above, and those
higher-level states that involve excitations {\it only} in the unfrozen
directions.  The consequence of this is that the spectrum of physical states
of the $W_N$ string is essentially given by the spectrum for Virasoro
strings with effective central charge (1.2), and a discrete set of
effective intercepts $L_0^{\rm eff}$.  We shall show that these values of
$L_0^{\rm eff}$ all satisfy the unitarity bounds arising from level-1 and
level-2 physical states.  These results provide a strong indication of the
unitarity of the $W_N$ string.

     The paper is organised as follows.  In the next section we review how
scalar realisations of $W_N$ may be obtained by using the Miura
transformation of $su(N)$, and we derive an explicit formula for $W_N$
currents in terms of $W_{N-1}$ currents and one extra scalar field (called
$\varphi_N$ in our notation).  Using these results, we prove the relation
between the central charges of the $W_N$ and $W_{N-1}$ algebras that was
conjectured in [5], and which leads to the relation between (1.1) and (1.2).
In section 3, we give the critical central charge for the $W_N$ string, and
determine the physical-state conditions.  In particular, this involves
calculating the higher-spin intercepts, which, {\it a priori},
one does not know without detailed knowledge of the BRST
operator for the $W_N$ gauge theory.  In [2], it was proposed that a
particular tachyonic operator called the ``cosmological-constant operator,''
for which the momentum is a certain multiple of the background-charge vector of
the $(N-1)$-scalar Miura realisation, is always a physical operator.  This
``cosmological solution'' enables one to determine the values of the
higher-spin intercepts, for which we give general formulae.
In section 4, we prove that the solutions to
the physical-state conditions for the tachyons of the $W_N$ string are
generated by the action of the Weyl group of $su(N)$ on
the cosmological solution.  We give a general proof for
arbitrary $W_N$ that this spectrum of states is associated with the
``diagonal'' highest-weight states for the $(N,N+1)$ Virasoro minimal model.
 In section 5, we consider higher-level physical states, and obtain
additional ``off-diagonal'' entries of the Kac table.  We also discuss the
no-ghost theorem for $W_N$ strings, and argue in particular that the
higher-level states with excitations in the frozen directions have zero norm
(these include all those associated with the off-diagonal entries of the Kac
table).  The remaining physical states have non-negative norm.  In
section 6, we return to the assumption of the existence of the
``cosmological-constant operator'' as a physical operator.  We give strong
arguments supporting this assumption, by demonstrating that if the values
of the higher-spin intercepts are different from those determined by this
particular physical operator, then one gets a non-unitary theory.  Finally,
in section 7, we present our conclusions and discuss some open problems.

\bigskip\bigskip
\noindent{\bf 2.  The Miura Transformation and the $W_N\rightarrow W_{N-1}$
Reduction.}
\bigskip

     A realisation of the $W_N$ algebra in terms of $(N-1)$ free scalars
$\vec\varphi^{(N)}\equiv(\varphi_2,\ldots,\varphi_N)$ is given by the Miura
transformation for $su(N)$ [4]
$$
\prod_{k=1}^N\Big(\a \del +\vec h^{(N)}_k \cdot (\del \vec\varphi^{(N)}) \Big)
=\sum_{\ell=0}^N W^{(N)}_\ell\, (\a \del)^{N-\ell},\eqno(2.1)
$$
where the $\vec h^{(N)}_k$ are $(N-1)$-component vectors satisfying
$$
\eqalign{
\vec h^{(N)}_i\cdot \vec h^{(N)}_j&=\delta_{ij}-{1\over N},\cr
\sum_{i=1}^N \vec h^{(N)}_i&=0.\cr}\eqno(2.2)
$$
It follows immediately from (2.1) that $W^{(N)}_0=1$ and $W^{(N)}_1=0$. The
quantities $W^{(N)}_\ell$ with $2\le \ell\le N$ are spin-$\ell$ currents
that generate the $W_N$ algebra.  They are not primary with respect to the
energy-momentum tensor $W^{(N)}_2$, but can be made so for $\ell\ge 3$ by
adding derivatives and composites of lower-spin currents.  Since this is not
essential for our purposes, we shall not take the trouble to do so.    By
convention, we shall always order products such as the one in (2.1) in
decreasing order of $k$, {\it i.e.}\ the largest-$k$ factor sits at the left.
Normal ordering of the quantum operators will always be understood.  The
fields $\varphi_i$ satisfy the operator-product expansions
$$
\varphi_i(z)\varphi_j(w)\sim -\delta_{ij}\log(z-w).\eqno(2.3)
$$

     The $\vec h^{(N)}_i$ for $1\le i\le N-1$ are the weights of the ${\bf
N}$ representation of $su(N)$, and $\vec h^{(N)}_N$ is defined by the second
equation in (2.2).  The simple roots $\vec e^{\,(N)}_i$ of
$su(N)$ are given in terms of these weights by
$$
\vec e^{\,(N)}_i=\vec h^{(N)}_i-\vec h^{(N)}_{i+1},\qquad 1\le i\le N-1.
\eqno(2.4)
$$
For later purposes we also introduce the Weyl vector $\vec \rho^{\,(N)}$,
given by
$$
\vec\rho^{\,(N)}=\sum_{j=1}^{N-1}\, (N-j)\vec h^{(N)}_j = \ft12\sum_{j=1}^{N-1}
j(N-j) \vec e^{\,(N)}_j\, .\eqno(2.5)
$$

     A convenient choice of representation for the weights $\vec h^{(N)}_i$ is
given by [2]
$$
\eqalign{
\vec h^{(N)}_1&=\Big({1\over \sqrt2},{1\over \sqrt6},{1\over \sqrt{12}},\ldots,
{1\over \sqrt{N(N-1)}}\Big),\cr
\vec h^{(N)}_p&=\Big(\underbrace{0,\ldots,0}_{p-2},-{p-1\over \sqrt{p(p-1)}},
{1\over \sqrt{p(p+1)}},\ldots,{1\over \sqrt{N(N-1)}}\Big),\cr}\eqno(2.6)
$$
where $p$ runs from 2 to $N$.

     This choice of vectors $\vec h^{(N)}_i$ for $W_N$ has the nice
property that
$$
\vec h^{(N)}_i=\Big(\vec h^{(N-1)}_i,{1\over \sqrt{N(N-1)}}\Big)\eqno(2.7)
$$
for $1\le i\le N-1$.  In other words, the first $(N-2)$ components of the
first $(N-1)$ vectors are precisely the $\vec h^{(N-1)}_i$ vectors for
$W_{N-1}$.  This enables one to re-express the $W_N$ currents in terms of
$W_{N-1}$ currents together with the scalar field $\varphi_N$ which is the
last component of the $\vec\varphi^{(N)}$ fields of the $W_N$ realisation.
To see this, we begin by writing the left-hand side of (2.1), using (2.7)
and (2.6), as
$$
\Big(\a\del -(N-1)(\del\phi_N)\Big)
\prod_{\ell=1}^{N-1}\Big(\a \del +\vec h^{(N-1)}_\ell \cdot (\del
\vec\varphi^{(N-1)}) +(\del\phi_N)\Big),\eqno(2.8)
$$
where we have defined
$$
\phi_N\equiv {1\over\sqrt{N(N-1)}}\varphi_N.\eqno(2.9)
$$
The $\prod_{\ell=1}^{N-1}$ factors in (2.8) may be rewritten as
$$
e^{-\phi_N/\a}\prod_{\ell=1}^{N-1}\Big(\a \del +\vec h^{(N-1)}_\ell \cdot
(\del \vec\varphi^{(N-1)}) \Big)e^{\phi_N/\a}.\eqno(2.10)
$$
Using the Miura transformation (2.1) for $W_{N-1}$, this can be written,
using Leibniz's rule, as
$$
\sum_{m=0}^{N-1}\,\,\sum_{q=0}^{N-m-1}{N-m-1\choose q}W^{(N-1)}_m P_q(\phi_N)
(\a\del)^{N-m-q-1},\eqno(2.11)
$$
where we have defined $P_q(\phi_N)$, which is a differential polynomial in
$\del\phi_N$, by
$$
P_q(\phi_N)\equiv e^{-\phi_N/\a}\Big( (\a\del)^q e^{\phi_N/\a}\Big).
\eqno(2.12)
$$
Note that $P_q(\phi_N)$ satisfies the recursion relation
$$
P_q(\phi_N)=\a\del P_{q-1}(\phi_N)+ \del \phi_N P_{q-1}(\phi_N).\eqno(2.13)
$$
Rewriting the double sum in (2.11), substituting back into (2.8), and equating
powers  of $(\a\del)$ in (2.1), we obtain the explicit relation
$$
\eqalign{
W^{(N)}_k=\sum_{q=0}^k &{N+q-k\choose q}\Big[ {N-k\over N+q-k}
W^{(N-1)}_{k-q} P_q(\phi_N)\cr
&+\a\del\Big(W^{(N-1)}_{k-q-1} P_q(\phi_N)\Big) -(N-1)(\del\phi_N)
W^{(N-1)}_{k-q-1} P_q(\phi_N)\Big].\cr}\eqno(2.14)
$$
All products of operators are understood to be normal ordered with respect
to the basic scalar fields $\varphi_i$.  Currents $W^{(N-1)}_m$ with $m<0$
are defined to be zero. (A formula equivalent to (2.10) was also derived in
[2].)

Equation (2.14) gives a realisation of the $W_N$ currents in terms of  those
for $W_{N-1}$, together with an additional scalar field $\varphi_N$.   Applying
this recursively leads to a realisation of the $W_N$ algebra in  terms of
$\varphi_2$, which appears only {\it via}\ its energy-momentum  tensor, and
$(N-2)$ additional scalar fields $(\varphi_3,\ldots,\varphi_N)$.  Since
$\varphi_2$ commutes with the other scalars, its energy-momentum  tensor may be
replaced by an arbitrary one that commutes with $(\varphi_3,\ldots,\varphi_N)$
and that has the same central charge.\footnote{$^\star$}{\tenfoot Note that
this allows one in particular  to realise $W_N$ on any affine Lie algebra $g$
with rank($g$) $\ge N-2$, generalising the results of [6].  The energy-momentum
tensor is then realised  on the coset $g/h$, where $h$ is an
$(N-2)$-dimensional Abelian subalgebra of  $g$, and the currents
$(\del\varphi_3,\ldots,\del\varphi_N)$ are taken to be  the generators of this
subalgebra.}

     Using equation (2.14) for $k=2$, we may relate the energy-momentum
tensors for $W_N$ and $W_{N-1}$ as follows:
$$
W^{(N)}_2=W^{(N-1)}_2 -\ft12(\del\varphi_N)^2+\ft12 \sqrt{N(N-1)}\, \a \del^2
\varphi_N.\eqno(2.15)
$$
Applying this recursively, we obtain
$$
\eqalign{
W^{(N)}_2&=\sum_{j=2}^N\Big( -\ft12(\del\varphi_j)^2 +\ft12\sqrt{j(j-1)}\, \a
\del^2\varphi_j\Big)\cr
&=-\ft12\del\vec\varphi^{(N)}\cdot \del\vec\varphi^{(N)} +\a
\vec\rho^{(N)}\cdot \del^2 \vec\varphi^{(N)},\cr}\eqno(2.16)
$$
which therefore generates the Virasoro algebra with central charge
$$
c_N=(N-1)\Big(1+N(N+1)\a^2\Big),\eqno(2.17)
$$
in agreement with [4].  The recursion relation
$$
c_N=-2+{N+1\over N-2}c_{N-1}\eqno(2.18)
$$
conjectured in [5] follows straightforwardly from (2.17), and the fact that
the background-charge parameter $\a$ for the $W_{N-1}$ Miura
transformation in (2.10) is identical to that for the $W_{N}$ transformation
(2.1).

     The scalar $\varphi_2$ appears in (2.16) (and indeed, as already
mentioned, in all the currents in (2.14)) {\it via}\ its energy-momentum
tensor
$$
T(\varphi_2)=-\ft12(\del\varphi_2)^2+{1\over \sqrt2} \a\del^2\varphi_2.
\eqno(2.19)
$$
This generates the Virasoro algebra with central charge $c=1+6\a^2$.  The
contribution from $\varphi_2$ can thus be replaced by an arbitrary
energy-momentum tensor with this central charge, {\it i.e.}\
$$
c={6c_N\over N(N^2-1)}+{2(N-2)(N+3)\over N(N+1)} -\Big(1-
{6\over N(N+1)}\Big).\eqno(2.20)
$$

\bigskip\bigskip
\noindent{\bf 3. Physical-state Conditions}
\bigskip

     We now turn to the consideration of the physical spectrum of the $W_N$
string. In the remaining sections of the paper, when there is no possibility of
confusion, we shall suppress the label $N$ that we have been using to indicate
that the quantities are associated with the $W_N$ algebra.

     Physical states $\phys$ of the $W_N$ string are defined by the conditions
$$
\eqalignno{
\big(W_s\big)_m \phys&=0,\qquad m\ge 1,&(3.1a)\cr
\big(W_s\big)_0 \phys&=\omega_s \phys,&(3.1b)\cr}
$$
where the Laurent modes $\big(W_s\big)_m$ of the spin-$s$ current
$W_s$ for the $W_N$ algebra are defined by $W_s(z)=\sum_m
\big(W_s\big)_m z^{-m-s}$.  The constants $\omega_s$ are the
intercepts for the zero modes of the spin-$s$ currents.  In principle, they
can be determined by requiring that the nilpotent BRST operator for the algebra
annihilate the physical vacuum (including ghosts).  In practice, however,
the construction of the BRST operator for the $W_N$ algebra is very
complicated, and has been given only for the case of $N= 3$ [7].  We shall
show later in this section how the intercepts may be determined by simpler
methods.  Note, incidentally, that it is sufficient to impose (3.1$b$) for
all $s$, together with (3.1$a$) for $s=2$ and $m=1$ and $m=2$, since the
rest of the constraints in (3.1$a$) then follow from the commutation
relations of the $W_N$ algebra.

     The requirement of nilpotency of the BRST operator determines the
central charge of the $W_N$-string theory.  Even though the $W_N$ algebra is
non-linear, the spin-2 current generates a linear subalgebra.  Thus the
total central charge, which must be zero for nilpotence, is simply the sum
of those for the matter and ghost sectors.  The ghosts for the spin-$s$
current contribute $-2(6s^2-6s+1)$ to the ghostly central charge, and so
the critical central charge $c^*_N$ for the matter sector is given by
$$
\eqalignno{
c^*_N&=2\sum_{s=2}^N (6s^2-6s+1)\cr
&=2(N-1)(2N^2+2N+1).&(3.2)\cr}
$$
{}From (2.17), we see that the background-charge parameter $\a$ is then given
by its critical value $\acrit$, namely
$$
(\acrit )^2={(2N+1)^2\over N(N+1)}.\eqno(3.3)
$$
{}From now on, we shall always assume that the central charge and $\a$ take
their critical values.

     The necessity of using BRST methods to determine the intercepts
$\omega_s$ in (3.1$b$) can be avoided if one knows a specific example
of an operator that creates a physical state, since then one can simply act
on it with $\big(W_s\big)_0$ and read off the values of the intercepts.  In
[2], such an operator, called the ``cosmological-constant operator'' was
proposed. Specifically, for $W_N$, it is a tachyonic operator of the form
$$
V_{{\rm cosmo}}=e^{\lambda \vec\rho\cdot
\vec\varphi},\eqno(3.4)
$$
where $\vec\rho$ is the Weyl vector given in (2.5), and
$\lambda$ is a certain constant to be determined.  For $W_3$, since
one knows the values of the intercepts from the BRST construction in [7], one
can explicitly verify that such a physical operator exists.  In [2], it was
argued from classical correspondence-principle considerations that such a
physical operator should occur for all higher $W_N$ algebras too.  In
section 6, we shall present a stronger argument that supports this
proposal.  For now, we shall proceed on the assumption that a physical
operator of the form (3.4) indeed exists.  It then remains to determine the
value of the constant $\lambda$.  This can be done by using an
independent argument that enables us to calculate the spin-2 intercept (see
for example [2,8]). Since $T^{{\rm tot}}\equiv T^{{\rm mat}}+T^{{\rm ghost}}$
annihilates $\phys \otimes \big|{\rm vac}\big\rangle_{{\rm ghost}}$, and the
spin-2 subalgebra of $W_N$ is linear,  we may read off the spin-2 intercept
as the negative of the intercept for the spin-2 ghost current acting on $
\big|{\rm vac}\big \rangle_{{\rm ghost}}$.  In other words, the BRST charge
is required, as usual, to annihilate $\phys \otimes \big|{\rm
vac}\big\rangle_{{\rm ghost}}$. The ghost vacuum is defined by
$$
\big|\hbox{vac}\big\rangle_{{\rm ghost}}\equiv\prod_{s=2}^N\prod_{m=1}^{s-1}
\big(c_s\big)_m \big|0\big\rangle,\eqno(3.5)
$$
where $\big|0\big\rangle$ is the $SL(2,C)$-invariant vacuum, and
$\big(c_s\big)_m$ are the Laurent modes of the usual ghost field for the
spin-$s$ current.  Thus we find that the spin-2 intercept is given by [2]
$$
\omega_2=\sum_{s=2}^N\, \sum_{m=1}^{s-1}\, m=\ft16 N(N^2-1).\eqno(3.6)
$$
{}From (2.16) and (3.4) it follows that $\lambda$ is given by
$$
\lambda=\Big(1\pm{1\over 2N+1}\Big)\acrit.\eqno(3.7)
$$
The two values for  $\lambda$ in (3.7) are related by a reflection symmetry,
as we shall explain in section 4.  Without  loss of generality, we shall
take the $+$ sign in (3.7), and refer  to the corresponding operator (3.4)
as the ``cosmological solution.''

We are now in a position to compute the intercepts for the $W_N$ string. To do
this, we first compute the eigenvalues of the zero modes of the $W_N$
currents acting on arbitrary tachyonic states, and then substitute the
cosmological solution defined in (3.4) and (3.7) into these eigenvalues.
Thus consider an arbitrary tachyonic operator
$$
V_{\vec\beta} = e^{\vec \beta\cdot \vec \varphi}. \eqno(3.8)
$$
The highest-order pole of the
operator-product expansion $W_s(z)  V_{\vec\beta}(w)$ is of order $s$, implying
that the tachyonic state, obtained from (3.8), satisfies (3.1$a$).  The
eigenvalue of this state under the action of  $\big(W_s\big)_0$ can be read
off from this highest-order pole. Since (3.8) satisfies
$$
\del \varphi_j(z)V_{\vec\beta}(0)\sim -{\beta_j V_{\vec\beta}(0)\over z}\ ,
\eqno(3.9)
$$
this pole can be obtained simply by replacing $\del \vec\varphi$ by
$-\vec\beta/z$ in formula (2.14) [4]. Let us define the functions
$U^{(n)}_s(z)$ for $2\le n\le N$ and $0\le s \le n$ by $U^{(n)}_0(z)=1$,
$U^{(n)}_1(z)=0$, $U^{(n)}_{t<0}(z)=0$ and the recursion relation
$$
\eqalign{
U^{(n)}_s(z)=\sum_{q=0}^s &{n+q-s\choose q}\Big[ {n-s\over n+q-s}
U^{(n-1)}_{s-q}(z) P_q(\zeta_n/z)\cr
&+\a\del\Big(U^{(n-1)}_{s-q-1}(z) P_q(\zeta_n/ z)\Big)  -(n-1){\zeta_n\over z}
U^{(n-1)}_{s-q-1}(z) P_q(\zeta_n/z)\Big],
\cr}\eqno(3.10)
$$
where we have introduced
$$
\zeta_n=-{\beta_n\over \sqrt{n(n-1)}}.\eqno(3.11)
$$
The eigenvalues $v_s(\vec\beta)$ of $V_{\vec\beta}$ under $\big(W_s\big)_0$
for the $W_N$ string are then given by
$$
v_s(\vec\beta)=U_s^{(N)}(z)\Big|_{z=1}.\eqno(3.12)
$$

   The intercepts $\omega_s$ for the $W_N$-string theory can be obtained by
substituting the cosmological solution (3.4), with $\lambda$ given by (3.7),
into (3.12) and (3.10), and solving the recursion relations. Even though we
have not been able to find a closed-form expression for all the $\omega_s$ as a
function of $N$ and $s$, we have found that the intercepts for any
$W_N$-string theory for some low spins $s$ can be written as
$$
\eqalign{
\omega_2 &= \ft16 (N+1)N(N-1),\cr
\omega_3 &= -\ft16 (N+1)N(N-1)(N-2)\,\acrit ,    \cr
\omega_4 &= \ft1{360} (N-1)(N-2)(N-3)(5N^3+228N^2+223N + 54), \cr
\omega_5 &=-\ft1{180} (N-1)(N-2)(N-3)(N-4)(5N^3+108N^2+103N+24)\, \acrit ,
 \cr
\omega_6 &= {1\over 45360 N(N+1)} (N-1)(N-2)(N-3)(N-4)(N-5)\cr
               & \times (35 N^6+ 7238N^5+110728N^4+ 201646 N^3 + 14378 N^2 +
                  45666 N + 5400).
\cr}\eqno(3.13)
$$
The expression for $\omega_2$ is just (3.6).  We shall show in section 6
how $\omega_3$ may be calculated in general.  For $s\ge4$, we have arrived
at the above expressions for $\omega_s$ by polynomial fitting from specific
results for small values of $N$; we have then verified these formulae for
all algebras up to and including $W_{20}$.

     An important remark is in order here. The intercepts computed above are
the ones for the $W_N$ currents as defined by the Miura transformation (2.1).
The intercepts for the primary $W_N$ currents can be obtained from the ones
presented here. As an example, the primary spin-3 current (for any $N$) is
given by $W_3 - \ft12 (N-2)\acrit \del W_2$ and hence has  intercept
$\omega_3 + (N-2)\acrit\,\omega_2 =0$, which agrees with the known result for
$N=3$ [7]. In general the intercepts for the primary currents are much more
complicated than the ones above. The determination of the physical states is a
basis-independent question and thus there is no advantage in working in the
more cumbersome ``primary basis.''  To avoid unnecessary complication we shall
therefore continue to use the more natural ``Miura basis.''

\bigskip\bigskip
\noindent{\bf 4. The Physical Spectrum for Tachyonic States}
\bigskip

   Having determined the intercepts for the $W_N$-string theory in the previous
section, we are now in a position to discuss the physical spectrum of this
theory. In this section we shall only consider tachyonic states, relegating
higher-level states to section 5.

\bigskip
\noindent{\it 4.1 The Weyl Reflection Symmetry of Tachyonic States}
\bigskip

     The eigenvalues  $v_s(\vec\beta)$  for the zero modes of $W_s$ acting on
the tachyonic state (3.8) can also be obtained directly from the Miura
transformation (2.1),  by making use of (3.9). One finds by letting the
differential operators act on $z^j$, for $0\le j\le N-2$, that [4]
$$
\sum_{s=0}^j {j! \over (j-s)!}(\acrit) ^s v_{N-s}(\vec\beta) =
\prod_{k=1}^N \Big[ \acrit (j+1-k) - \vec h_k \cdot \vec\beta \Big].
\eqno(4.1)
$$
Shifting $\vec\beta$, so that
$$
\vec\beta=\vec\gamma + \acrit\,\vec\rho,\eqno(4.2)
$$
leads to
$$
\sum_{s=0}^j {j! \over (j-s)!}(\acrit) ^s v_{N-s}(\vec\beta) =
\prod_{k=1}^N \Big[\ft12 \acrit (2j+1-N) - \vec h_k \cdot \vec\gamma \Big].
\eqno(4.3)
$$
{}From this identity one can see that the eigenvalues $v_s(\vec\beta)$, which
are now polynomials in the shifted momentum $\vec \gamma$,
are invariant under $su(N)$ Weyl reflections of $\vec\gamma$. To show this,
we first note that Weyl reflections of the simple roots of
$su(N)$, as given in (2.4), are defined by
$$
S_{\vec e_i}(\vec e_j)\equiv \vec e_j - (\vec e_j\cdot \vec e_i)
\vec e_i. \eqno(4.4)
$$
This implies that the action of $S_{\vec e_i}$ on $\vec h_k$
interchanges  $\vec h_i$ with $\vec h_{i+1}$ whilst leaving all the other $\vec
h_k$ fixed.  From the invariance of the scalar product, it follows that a Weyl
reflection of $\vec\gamma$
$$
\vec\gamma \longrightarrow S_{\vec e_i}(\vec\gamma)=\vec \gamma-
(\vec\gamma\cdot\vec e_i) \vec e_i, \eqno(4.5)
$$
for any simple root $\vec e_i$ of $su(N)$, merely permutes the ordering of
the factors in the right-hand side of (4.3).  Since the simple roots
generate the entire Weyl group, we conclude that indeed the polynomials $v_s$
are invariant under all Weyl reflections of the shifted momentum $\vec\gamma$.

\np

\noindent{\it 4.2 The Tachyon Spectrum of $W_N$-String Theory}
\bigskip

     To determine the tachyonic states in the $W_N$-string
spectrum, one has to solve the physical-state conditions (3.1$b$), {\it
i.e.}
$$
v_s(\vec \beta)=\omega_s,\qquad 2\le s\le N,\eqno(4.6)
$$
where the intercepts $\omega_s$ are given by $v_s(\lambda\vec\rho)$,
according to the discussion of the previous section.  Since $v_s(\vec\beta)$
is a polynomial of degree $s$ in $\vec\beta$, it follows that the set of
equations (4.6) will have $N!$ solutions.  Because, as we have seen, the Weyl
group of $su(N)$ acting on the shifted momentum $\vec\gamma$ leaves (4.6)
invariant, we learn that it maps solutions of (4.6) into solutions.  In
fact, we know one solution of (4.6), {\it viz.}\ the cosmological  solution,
which is given by (3.4) and (3.7); this is a solution by construction.  Since
the Weyl vector $\vec\rho$ is not a fixed point of the Weyl group, and the
dimension of the Weyl group of $su(N)$ is $N!$, we therefore conclude that we
obtain {\it all} the tachyonic physical states  from the cosmological
solution  by the action of the Weyl group on it.

     An explicit procedure for writing down the $N!$ elements of the Weyl
group of $su(N)$ can be described as follows.  Defining $S_i\equiv S_{\vec
e_i}$, these elements can be obtained by taking the product of one entry
from each column of
$$
\left\lgroup\matrix{1\cr S_1\cr}\right\rgroup\otimes
\left\lgroup\matrix{1\cr S_2\cr S_2 S_1\cr}\right\rgroup\otimes
\left\lgroup\matrix{1\cr S_3\cr S_3 S_2\cr S_3 S_2 S_1\cr}\right\rgroup\otimes
\cdots\otimes
\left\lgroup\matrix{1\cr S_{N-1}\cr S_{N-1}S_{N-2}\cr\vdots\cr
S_{N-1}S_{N-2}\cdots S_1\cr}\right\rgroup,\eqno(4.7)
$$
giving $N!$ (inequivalent) choices in all.  Applying these to the
shifted momentum
$$
\vec\gamma^{\rm cosmo}={\acrit\over 2N+1}\,\vec\rho\eqno(4.8)
$$
of the cosmological solution fills out all the $N!$ tachyonic physical
states of the $W_N$ string.  Note that included amongst these $N!$ solutions
generated by (4.7) is one that corresponds to taking the $-$ sign instead of
the $+$ sign in (3.7).  It is obtained by choosing the Weyl reflection
generated by the product of the bottom entries in each column of (4.7), and
corresponds to the reflection $\vec\rho\rightarrow -\vec\rho$.

\bigskip
\noindent{\it 4.3 The Target Spacetime of $W_N$ Strings}
\bigskip

     In the previous subsection, we showed how the set of $N!$ tachyonic
physical states of the $W_N$ string are generated by the Weyl group acting
on the cosmological solution given by (3.4) and (3.7).  In the discussions
so far, we have considered an $(N-1)$-dimensional target space, with
coordinates $(\varphi_2,\varphi_3,\ldots,\varphi_N)$.  Since the
physical-state conditions imply that the momentum components $\beta_j$ can
take only specific, discrete values (for example, the $N!$ tachyon
solutions), there is no sensible notion of a physical spacetime yet.  To
obtain a physical-spacetime interpretation, we can carry out the procedure
described in section 2, of replacing the energy-momentum tensor of the
$\varphi_2$ scalar by an arbitrary energy-momentum tensor with the same
central charge, which is obtained by substituting the critical value $c_N^*$
for $W_N$, given by (3.2), into (2.20), leading to the expression given
in (1.2).

     We shall take this energy-momentum tensor to be that for $\varphi_2$
plus $D$ additional free scalar fields $X^\mu$, one of which will be chosen
to be timelike, and the rest spacelike.  Thus we have
$$
T^{\rm eff}=-\ft12\big(\del\varphi_2\big)^2 +Q \del^2\varphi_2
-\ft12 \eta_{\mu\nu}\del X^\mu\, \del X^\nu.\eqno(4.9)
$$
The background charge $Q$ must be chosen so that $T^{\rm eff}$ has central
charge given by (1.2), and so
$$
\eqalign{
Q^2&=\ft1{12}\Big(6(\acrit)^2-D\Big)\cr
&=\ft1{12}\Big(24+{6\over N(N+1)}-D\Big).\cr}\eqno(4.10)
$$
Note that $Q$ is non-zero for all $W_N$ strings (with $N\ge3$), regardless
of the number $D$ of additional scalars $X^\mu$.  It is for this reason that
we choose to separate the coordinates into $\varphi_2$, which carries the
background charge, and the remaining $X^\mu$, which have no background
charge.

     When one realises the $W_N$ algebra with just the $(N-1)$ scalars
$\vec\varphi$, all $(N-1)$ components of the momentum $\vec\beta$ are
``frozen'' by the physical-state conditions to specific discrete sets of
values $\vec\beta^{\rm froz}$, such as those of the $N!$ tachyonic
solutions which we are considering in this section.  The effect of
introducing extra scalar fields $X^\mu$ is that the momentum components
$(\beta_3,\beta_4,\ldots,\beta_N)$ continue to be frozen to exactly the same
sets of values $(\beta_3^{\rm froz},\beta_4^{\rm froz},\ldots,\beta_N^{\rm
froz})$, whilst the momentum components $(\beta_2,\beta_\mu)$ satisfy
$$
L^{\rm eff}_0=
-\ft12\beta_2^2 +Q\beta_2 -\ft12 \beta_\mu \beta^\mu,\eqno(4.11)
$$
where
$$
L^{\rm eff}_0\equiv -\ft12\big(\beta_2^{\rm froz}\big)^2
+{1\over \sqrt2}\acrit\,\beta_2^{\rm froz}.\eqno(4.12)
$$
The fact that the momentum components $(\beta_3,\beta_4,\ldots,\beta_N)$
remain unchanged, and the remaining $\beta$'s satisfy (4.11), is a
consequence of the special way in which the new coordinates $X^\mu$ are
introduced into the theory, as a modification of the original
energy-momentum tensor for $\varphi_2$.  Thus $\beta_2^{\rm froz}$ appeared
in the physical-state conditions only through the combination $L^{\rm
eff}_0$ defined by (4.12), and so after adding the extra coordinates
$L^{\rm eff}_0$ remains unchanged.

     The conclusion of the above discussion is that we can effectively view
the tachyonic spectrum of the $W_N$ string as being composed of sets of
Virasoro-type physical states, all with the same central charge $c^{\rm
eff}$ given by (1.2), but with different intercepts $L^{\rm eff}_0$ given by
substituting the discrete $\beta_2^{\rm froz}$ values into (4.12).  In
section 5, we shall show that this in fact holds for the spectrum of
higher-level states also.

\bigskip
\noindent{\it 4.4 $W_N$ Strings and Minimal Models}
\bigskip

     In [2], it was noticed that the tachyonic physical states for the  $W_3$
and $W_4$ strings display a numerological connection with  the  Virasoro
minimal models with central charges $c=1/2$ and $c=7/10$  respectively.  In
this
subsection, we shall generalise and explain this  numerological connection.

     As already mentioned in the introduction, one can rewrite $c^{\rm  eff}$
in the suggestive form (1.3), which consists of a term equal to the critical
central charge for the usual Virasoro string minus the central charge of a
unitary minimal model.  This is indicative of a possible connection between the
$W_N$ string and the $(N,N+1)$ Virasoro minimal model with central charge
$$
c(N)=1-{6 \over N(N+1)}\ .\eqno(4.13)
$$
The suggestion is strengthened by the fact that if one rewrites $L^{\rm
eff}_0$ as
$$
L^{\rm eff}_0=1-L^{\rm min}_0\ ,\eqno(4.14)
$$
where 1 is the value for the intercept of the critical Virasoro string,  then
$L^{\rm min}_0$ is precisely the dimension of a primary field of the minimal
model with central charge given by (4.13).  The values of $L^{\rm eff}_0$
corresponding to all the tachyonic physical states can be obtained from (4.12).
Substituting these values  into (4.14)  yields all the ``diagonal'' entries of
the Kac table of the relevant minimal model, as we now show.

     Since the values of $L^{\rm eff}_0$ can be determined by
$\beta^{\rm froz}_2$, we need only compute these latter values. They are
easily obtained by acting with the Weyl group on the cosmological solution.
It turns out that for the $W_N$ string the shifted momentum components
$\gamma^{\rm froz}_2$ can take the following values
$$
\gamma^{\rm froz}_2 = \pm {\acrit\, k \over \sqrt{2}(2N+1)}, \eqno(4.15)
$$
where $k$ is an integer satisfying
$$
1\le k\le N-1. \eqno(4.16)
$$
One can easily see this from the Weyl reflections
$$
\left\lgroup\matrix{1\cr S_1\cr}\right\rgroup\otimes S_2\, S_3\,
\cdots S_k(\vec\rho)= \vec\rho-\sum_{j=(2,1)}^k (k-j+1)\vec e_j,\eqno(4.17)
$$
where the first lower bound in the summation of the right-hand side corresponds
to the upper entry in left-hand side, and the second bound to the lower entry.
{}From (4.2), (4.12), (4.14) and (4.15), we therefore find the following
values for $L_0^{\rm min}$:
$$
L_0^{\rm min} = {k^2-1\over4N(N+1)}, \eqno(4.18)
$$
where the integer $k$ lies in the interval given by (4.16). The dimensions of
the primary fields of the Virasoro minimal model with central charge given by
(4.13) are
$$
\Delta_{(r,s)}= {\big[ (N+1)r -Ns\big]^2-1\over4N(N+1)}, \eqno(4.19)
$$
where $r$ and $s$ are integers lying in the ranges $1\le r\le N-1$ and  $1\le s
\le N$ respectively. Thus we see that the weights in (4.18) precisely
correspond to the $s=r=k$ entries in the Kac table of the minimal model, {\it
i.e.}\ the ``diagonal'' ones.  In section 5, we shall show how other entries of
the Kac table arise from higher-level  physical states.

    We have exhibited $2(N-1)$ Weyl reflections (4.17) which generate the
$2(N-1)$ distinct values of $\beta_2^{\rm froz}$ of the tachyonic
states of the $W_N$ string. Since there are $N!$ such states, it follows
that in general $\beta_2^{\rm froz}$ is degenerate. Indeed, we find that
$(N-k) (N-2)!$ different tachyonic states have identical shifted-momentum
component $\gamma_2^{\rm froz}$  for each $k$ and each choice of sign in
(4.15). Thus for each allowed $k$, the value of $L_0^{\rm
min}=\Delta_{(k,k)}$ in (4.18) occurs with degeneracy $2(N-k)(N-2)!$.

     The relation between $W_N$ strings and the dimensions of the primary
fields of Virasoro minimal models that we have just described is, in fact, a
special case of a more general association that can be made between $W_N$
strings and the dimensions for minimal models of $W_M$ algebras, with $M<
N$. Instead of replacing just $\varphi_2$ by a new energy-momentum tensor,
one could as well replace $(\varphi_2,\ldots,\varphi_M)$ by new $W_M$
currents. These now have to satisfy the $W_M$ algebra with central charge
$$
c_M^{\rm eff}= c_M^*-c_M^{\rm min}(N), \eqno(4.20)
$$
where $c_M^*$ is the critical central charge for $W_M$ (generalising
the 26 of (1.3)) and the remainder,
$$
c_M^{\rm min}(N)=(M-1)\Big[1-{M(M+1)\over N(N+1)}\Big], \eqno(4.21)
$$
corresponds to the central charge of the relevant $W_M$ minimal model.
By analogy with (4.14), we write
$$
L_0^{\rm eff} (\beta_2,\beta_3,\ldots,\beta_M) = \ft16 M(M^2-1) -
L^{{\rm min}\,M}_0, \eqno(4.22)
$$
where $L_0^{\rm eff} (\beta_2,\beta_3,\ldots,\beta_M)$ is the contribution to
the spin-2 intercept of the $W_N$ string from the scalars realising $W_M$,
and is fixed by the physical-state conditions. By substituting all possible
values of the frozen-momentum components $(\beta_2^{\rm froz},\beta_3^{\rm
froz},\ldots, \beta_M^{\rm froz})$, obtained by acting with (4.7) on the
cosmological solution (4.8), into (4.22), one then finds that $L^{{\rm
min}\,M}_0$ takes values in the Kac table of the corresponding minimal model
of $W_M$. For example, if $M=N-1$, one finds the following values for
$L^{{\rm min}\,M}_0$:
$$
L^{{\rm min}\,M}_0= {k(N-k-1)\over 2(N^2-1)}, \qquad k=0,1,\ldots,
\Big[{N-1\over 2}\Big], \eqno(4.23)
$$
where $[x]$ denotes the integer part of $x$. The values given in (4.23) are
equal to the ``diagonal'' entries of the Kac table\footnote{$^\star$}{\tenfoot
By diagonal, we mean the dimensions with $\ell_i=\ell'_i$ in the notation of
[4,9].} of the $W_{N-1}$  minimal  model with central charge given by
$c_{N-1}^{\rm min}(N)$ as defined in (4.21).  The connection between
$W_N$ strings and $W_M$ minimal models is, however, not so clear as in the
case of Virasoro minimal models, since there does not seem to be any way of
defining effective intercepts for the higher-spin currents analogous to
$L_0^{\rm eff}$.

\bigskip\bigskip

\noindent {\bf 5. Higher-level Physical States and the No-ghost Theorem}
\bigskip

    Having discussed the tachyonic states in the previous section, we
shall now turn our attention to higher-level states.  If one considers
cases where the excitations occur purely in the unfrozen directions
$\varphi_2$ and $X^\mu$, the analysis is similar to that for ordinary string
theory, and can be carried out for arbitrary $W_N$ strings, at arbitrary
level number.  For excitations involving the frozen directions
$(\varphi_3,\ldots,\varphi_N)$, the analysis is much more complicated.  By
looking at special cases for the $W_3$, $W_4$, and $W_5$ strings, a general
pattern seems to emerge, indicating that these states have frozen-momentum
components that are incompatible with momentum conservation in their
two-point functions, and thus that they have zero norm.  Consequently, the
physical spectrum of higher-level states comprises excitations only in the
unfrozen directions.   For these we show,  by looking at level-1 and
level-2 states, that they have non-negative norms.

\bigskip
\noindent {\it 5.1 The Higher-level Physical Spectrum of $W_N$ Strings}
\bigskip

     The most general level-1 state in the $(N-1)$-scalar realisation of
the $W_N$ algebra is given by
$$
V_{(\vec\xi,\vec\beta)}(z)= \vec\xi\cdot \del\vec\varphi\, e^{\vec\beta\cdot
\vec\varphi},\eqno(5.1)
$$
where $\vec\xi\equiv (\xi_2,\xi_3,\ldots,\xi_N)$ is a polarisation vector.
In addition to the zero-mode constraints (3.1$b$), there is one other
non-trivial constraint coming from (3.1$a$):
$$
\big(W_2\big)_1 V_{(\vec\xi,\vec\beta)}(0)\big|0\big\rangle =0.
\eqno(5.2)
$$

     In the tachyonic case the physical-state conditions can be treated in
a general way, since these states are automatically eigenstates of the zero
modes $\big(W_s\big)_0$ of the $W_N$ currents, and since their eigenvalues
can be obtained by algebraic means; see (3.12). For level-1 states, the
physical-state conditions are much more complicated: level-1 states are not
automatically eigenstates of $\big(W_s\big)_0$; they have to satisfy the
additional condition (5.2); and finally, one has to compute non-trivial
operator-product expansions in order to find the explicit form of the
constraints.  For these reasons, we have not found a treatment as general as
the one for the tachyonic case. However we shall present a general pattern
for certain level-1 states, based on a complete analysis for $W_3$, $W_4$
and $W_5$, suggesting a generalisation to $W_N$ for arbitrary $N$. Moreover,
we shall also give the results of our complete analysis of the level-2
physical states for the $W_3$ and $W_4$ cases.

     Let us consider splitting the scalars $(\varphi_2,\ldots,\varphi_N)$ into
two sets: $\vec\varphi_\flat\equiv (\varphi_2,\ldots, \varphi_s)$ and
$\vec\varphi_\sharp\equiv (\varphi_{s+1},\ldots,\varphi_N)$, where $s$ is a
fixed integer between 2 and $N-1$. Consider a general physical operator of the
form
$$
P(\vec\varphi_\flat,\vec\varphi_\sharp)= \widetilde P(\vec\varphi_\flat) \,
e^{\vec\beta_\sharp\cdot
\vec\varphi_\sharp}, \eqno(5.3)
$$
where $\widetilde P(\vec\varphi_\flat)$ is of the form
$$
\widetilde P (\vec\varphi_\flat)= \widetilde R(\vec\varphi_\flat) \,
e^{\vec\beta_\flat\cdot \vec\varphi_\flat},\eqno(5.4)
$$
with $\widetilde R (\vec\varphi_\flat)$ a given differential polynomial in
$(\del\varphi_2,\ldots,\del\varphi_s)$.  This means that the corresponding
state has excitations only in the $(\varphi_2,\ldots,\varphi_s)$ directions.
One can then show that the values to which the momentum components
$(\beta_{s+1},\ldots,\beta_N)$ of the operator (5.3)  are frozen by the
physical-state conditions are independent of the detailed structure of
$\widetilde P(\vec\varphi_\flat)$ and are therefore equal to the
corresponding frozen-momentum components of the tachyonic physical
states.\footnote{$^\ddagger$}{\tenfoot This follows from the fact that the
$W_N$ currents can be written as linear combinations of $W^{(N-1)}_2,
\ldots,$ $W^{(N-1)}_s$ (which do not depend on $\vec\varphi_\sharp$) with
coefficients which are differential  polynomials in $\del\vec\varphi_\sharp$.
The detailed structure of  $\widetilde P(\vec\varphi_\flat)$ then enters the
physical-state conditions only through the operator-product expansions
between $W^{(N-1)}_j$, for $j=2,3, \ldots, s$, and $\widetilde
P(\vec\varphi_\flat)$, which can be rewritten, using (2.14), as
operator-product expansions between $W^{(N)}_j$ currents  and the physical
operator $P(\vec\varphi_\flat,\vec\varphi_\sharp)$. One can then replace
$W_j^{(N)}$ by its intercept value $\omega^{(N)}_j$, which is independent of
the structure of $\widetilde P(\vec\varphi_\flat)$.}

     An important result that follows from this property is that higher-level
physical states of the $W_N$-string theory (for all levels) defined by (5.3)
with $s=2$ will have the momentum components $(\beta_3, \ldots, \beta_N)$
frozen identically to their tachyonic values. Thus these physical states are,
just like the tachyonic physical states, related to the ``diagonal'' entries
in the Kac table of the $(N,N+1)$ Virasoro minimal model. The physical-state
conditions for states  which are not of this form are much more complicated
and we have not been able to solve them in general.  However, the complete set
of level-1 physical states for $W_3$, $W_4$ and $W_5$ suggests that the
higher-level states for any $W_N$-string theory will recover the remaining
entries of the Kac table.

    Let us first consider level-1 physical states of the form (5.1) for the
$W_3$-string theory. As the case $\vec\xi=(\xi_2,0)$ has already been
discussed, we shall assume that $\xi_3 \neq 0$. We need only give the
results for the momentum component $\beta_2$, since these are sufficient to
establish the connection with the corresponding minimal model. Solving the
physical-state conditions in this case leads to six possible values of
$\beta_2$. In terms of the shifted-momentum component $\gamma_2$ they are
given by (4.15), where now  $k\in \{1,2,5\}$.  Using equation (4.14), the $k=1$
and $k=2$ values correspond  to the diagonal entries $0$ and $1/16$ of the Kac
table of the Virasoro  minimal model with $c=1/2$, whereas $k=5$ gives the
remaining dimension of this minimal model, namely $\Delta_{(2,1)}=1/2$. Thus,
by
considering level-1 physical states of the $W_3$ string one finds all the
dimensions of the $c=1/2$ Virasoro minimal model.

    Consider now the $W_4$ string. Suppose first that the polarisation
components $\xi_3$ and $\xi_4$ are both non-zero. In that case, the
shifted-momentum component $\gamma_2$ is again given by (4.15), where now
$k\in\{1,2,3,6,7\}$.  The values $k=1$, $k=2$ and $k=3$ lead to the diagonal
entries of the $c=7/10$ Virasoro minimal models, whereas $k=6$ and $k=7$ lead
to the off-diagonal dimensions $\Delta_{(2,1)}=7/16$ and $\Delta_{(3,2)}= 3/5$
respectively. Only one dimension of this minimal model has not yet been found,
namely $\Delta_{(3,3)}=3/2$. The cases where $\xi_3=0$ or $\xi_4=0$ do not lead
to anything new.

    The $W_5$ case is again very similar. The values of $k$ which do not lead
to
diagonal entries of the $c=4/5$ Virasoro minimal model are now 7, 8 and 9.
They lead to the off-diagonal entries $\Delta_{(2,1)}=2/5$, $\Delta_{(3,2)}=
21/40$ and $\Delta_{(4,3)}=2/3$ respectively.

    The pattern that seems to emerge for the level-1 physical states of the
general  $W_N$ string should now be clear. The new values of $k$ in (4.15) and
(4.18) are given by
$$
k\in \{N+2,N+3, \ldots , 2N-1\}\eqno(5.5)
$$
and lead to the off-diagonal dimensions $\Delta_{(r,r-1)}$, with $r=2,
3,\ldots, N-1$, of the $(N,N+1)$ Virasoro minimal model.

    The analysis of the spectrum of level-1 physical states for the
$W_N$ string thus gives more evidence for the connection between the $W_N$
string and the $(N,N+1)$ Virasoro minimal model.  For $W_3$ this level-1
spectrum exhausts all the dimensions of the relevant minimal model. For the
other $W_N$-string theories we expect the remaining off-diagonal dimensions
to appear from physical states at sufficiently high level.

    We have, in fact, constructed all the level-2 physical states of the $W_3$
and $W_4$ strings. All these states are, once again, related to the relevant
minimal model.  For $W_4$, however, there is still no physical state at this
level which corresponds to the dimension $3/2$ primary field of the $c=7/10$
Virasoro minimal model. We expect, nevertheless, that this state will emerge
from the higher-level physical spectrum.

    In subsection 4.4 we generalised the connection between $W_N$ strings and
Virasoro minimal models in the tachyonic case to a connection with $W_M$
minimal models. This generalisation seems to hold for the higher-level physical
spectrum as well, where off-diagonal dimensions of the $W_M$ minimal models
appear. We have checked this explicitly for the connection between the
$W_4$-string theory and the $W_3$ minimal model with $c=4/5$. In this case the
two off-diagonal entries, $2/3$ and $2/5$, of the corresponding Kac table, as
well as the diagonal ones, $0$ and $1/15$, emerge in the level-1 and level-2
physical spectrum of the $W_4$ string.

\bigskip
\noindent {\it 5.2 The No-ghost Theorem for $W_N$ Strings}
\bigskip

     Having analysed the physical spectrum of the $W_N$ string, we are now
going to use these results to discuss the no-ghost theorem for $W_N$-string
theories.   As explained in subsection 4.3, the target spacetime of the
$W_N$ string only acquires a physical interpretation if the energy-momentum
tensor for $\varphi_2$ is replaced by a new energy-momentum tensor (4.9)
with central charge $c^{\rm eff}$ given in (1.2). We shall first show that
after adding extra coordinates $X^\mu$, all higher-level physical states
that involve excitations only in the unfrozen directions, {\it i.e.}\ that
are of the form (5.3) with $s=2$ (and $\varphi_2$ replaced by the set $\{
\varphi_2, X^\mu\}$), have positive semi-definite norm.  Next, we shall
argue, based on some explicit examples, that all the physical states that
are not of this form have zero norm and hence do not describe physical
degrees of freedom.  This absence of negative-norm states at low-lying
levels is usually a good indication of the ghost freedom of the theory.

     Higher-level physical operators of the form
$$
R(\varphi_2,X^\mu)\, e^{\vec\beta\cdot \vec\varphi + \beta_\mu X^\mu}
\eqno(5.6)
$$
have, as we have explained in the previous subsection, the same set of
frozen values for the momentum components $(\beta_3^{\rm
froz},\ldots,\beta_N^{\rm froz})$, regardless of the explicit form of
$R(\varphi_2,X^\mu)$, {\it i.e.}\ of their level. This implies that all the
states of the form (5.6) with given values of $(\beta_3^{\rm
froz},\ldots,\beta_N^{\rm froz})$ have the same value of $L_0^{\rm eff}$:
$$
\eqalign{
L_0^{\rm eff}&\equiv \ft16N(N^2-1)-\sum_{j=3}^N \Big[ -\ft12 (\beta_j^{\rm
froz})^2 + {1\over \sqrt{j(j-1)}}\acrit\, \beta_j^{\rm froz}\Big]\cr
&= n -\ft12 \beta_2^2+ Q\beta_2-\ft12 \beta_\mu\beta^\mu,\cr} \eqno(5.7)
$$
where $n$ is the level number of the states (5.6), and the background charge
$Q$ is given by (4.10). The computation of the norm of such physical states is
therefore analogous to that for physical states in Virasoro-string theory, but
with central charge $c^{\rm eff}$ and intercept $L_0^{\rm eff}$,  given in
(1.2) and (5.7). One can thus use the well-known method of deriving  unitarity
bounds for the intercept $a$ in Virasoro-string theory with a given central
charge $c$.  For level-1 physical states this bound is independent of the
value of the central charge and is given by $a \le 1$. For level 2, the bounds
depend on the value of the central charge:  they are given by
$$
a\le {37-c-\sqrt{(c-1)(c-25)}\over 16} \qquad {\rm or} \qquad
a\ge {37-c+\sqrt{(c-1)(c-25)}\over 16}.\eqno(5.8)
$$
In string theory, these bounds are sufficient to establish the unitarity of
the theory at all levels.  Combining these bounds for the central charge
$c^{\rm eff}$ given by (1.2) requires the intercept to satisfy
$$
a\le {3(N-1)\over4N}  \qquad {\rm or} \qquad  {3(N+2)\over4(N+1)} \le a \le
1.\eqno(5.9)
$$

     The values of $L_0^{\rm eff}$ given in (5.7) can be most easily obtained
by using the tachyonic physical states. Substituting (4.18) into (4.14) gives
the values
$$
L_0^{\rm eff}= {(2N+1)^2-k^2\over 4N(N+1)},\qquad k=1,2,\ldots, N-1.
\eqno(5.10)
$$
Each value of $L_0^{\rm eff}$ in (5.10) corresponds to the value of the
intercept for a Virasoro-type string with central charge (1.2). One can
easily see that each such intercept satisfies the unitarity bounds (5.9). We
therefore conclude that all the physical states of the form (5.6) have
positive semi-definite norm.  This demonstration that physical states having
excitations only in the unfrozen directions have non-negative norm concludes
the first part of the no-ghost theorem.

     The above discussion is not valid any more for states that are not of the
form (5.6). However, we shall argue, based on momentum-conservation
considerations, that all these states have vanishing norm.  The point is that
higher-level physical states are always of the form $\big|{\rm
phys}\big\rangle= R\big|p\big\rangle$, where $\big|p\big\rangle$ is a
tachyon-like state and  $R$ is a differential polynomial in the free scalars
with polarisation tensors as coefficients. The norms of such states are
$$
\big\langle{\rm phys}\big|{\rm phys}\big\rangle = {\cal N}(R)
\big\langle p \big| p \big\rangle\ , \eqno(5.11)
$$
where ${\cal N}(R)$ is a function of the scalar products of the polarisation
tensors.  The physical state $\big|{\rm phys}\big\rangle$ is a null state if
$\big\langle p \big| p \big\rangle=0$, which occurs when momentum conservation
cannot be satisfied.  Under these circumstances, the sign of ${\cal N}(R)$
is immaterial.

     The momentum-conservation law for tachyonic states appears as a delta
function in their two-point function.  For a $W_N$-string theory with effective
energy-momentum tensor given in equation (4.9) this is expressed by
\np
$$
\Big\langle e^{\vec\beta'\cdot\vec\varphi+ \beta'_\mu X^\mu}
e^{\vec\beta\cdot\vec\varphi + \beta_\mu X^\mu} \Big\rangle
\,\propto\, \delta(\beta'_2+\beta_2 -2 Q ) \prod_\mu\,
\delta(\beta'_\mu+\beta_\mu) \prod_{j=3}^N\, \delta (\beta'_j +\beta_j -2
\acrit \rho_j)\ , \eqno(5.12)
$$
where $\rho_j=\sqrt{j(j-1)}/2$ is the $j$-th component of the Weyl vector.
Since the  addition of extra coordinates allows $\beta_2$ and $\beta_\mu$ to
take continuous values, while leaving $(\beta_3,\ldots, \beta_N)$ frozen to
the same discrete values, it may happen that the momentum-conservation law
cannot be satisfied in the $(\varphi_3,\ldots,\varphi_N)$ directions. In
this case the delta function is zero in these directions, implying that the
two-point function (5.12) vanishes.

     Let us first consider the tachyonic states. Since, in order to discuss
their physical spectrum, it is not essential to introduce extra coordinates
$X^\mu$, we shall restrict ourselves to the $(N-1)$-scalar realisation. Suppose
that $\vec\beta_+=\acrit\,\vec\rho +\vec\gamma$ is a solution of the
physical-state conditions for the tachyon case (which means  that $\vec\gamma$
can be obtained from a Weyl reflection (4.7) on $\vec\gamma^{\rm cosmo}$ given
in (4.8)).  It then follows that $\vec\beta_-= \acrit\,\vec\rho -\vec\gamma$ is
also a solution of the physical-state conditions.  Since $\vec\beta_+ +
\vec\beta_-=2\acrit\, \vec\rho$, we see from (5.12) that the momenta
$\vec\beta_+$ and $\vec\beta_-$ are conjugate, implying that the
corresponding two-point function (5.12) is not zero. Thus, tachyonic states
have positive norm. Since physical operators of the form (5.6) have their
momentum components $(\beta_3,\ldots,\beta_N)$ frozen identically to the
tachyonic values, the same argument teaches us that these states are, in
general, not null states.

     The conclusion for higher-level physical states that have excitations
in frozen directions is different. In this case, the frozen momenta
$\vec\beta^{\rm froz}$ do not seem to appear in conjugate pairs, and
therefore the momentum-conservation law cannot be satisfied. Any two-point
function between such states is thus zero, implying that all these states
are null. Although we do not have a general proof of this statement, we have
checked it for all level-1 states of the $W_3$, $W_4$ and $W_5$ strings and
for all level-2 states of the $W_3$ and $W_4$ strings.  Indeed we find in
all these examples that the component $\gamma_s$ ($s\ge 3$) of the shifted
momentum $\vec\gamma$ is always positive, where $s$ is defined just above
equation (5.3). The condition for having a conjugate-momentum pair
$\vec\beta$ and $\vec\beta'$ is, as we have seen above, that their shifted
momenta $\vec\gamma$ and $\vec\gamma'$ should satisfy
$\vec\gamma=-\vec\gamma'$.  Since $\gamma_s$ is always positive for the
states in question, such conjugate pairs do not occur.  It seems reasonable
to expect that this pattern will hold in general.  Proving that all states
involving excitations in the frozen directions have zero norm, and thus do
not contribute to the physical spectrum, establishes the second part of the
no-ghost theorem for $W_N$ strings.

     Summarising this discussion, we conclude that the complete physical
spectrum of the $W_N$ string is given by the tachyonic states discussed in
section 4, together with all higher-level physical states that have
excitations only in the unfrozen directions.

\np

\noindent {\bf 6. Uniqueness of the Higher-spin Intercepts}
\bigskip

      In section 3 we computed the intercepts for the $W_N$ string under the
assumption that a particular tachyonic operator, namely the cosmological
operator (3.4), is physical.  After determining the constant $\lambda$ in
(3.4) by using the known spin-2 intercept (3.6), the intercepts for all
higher-spin currents followed.

     In [2], a classical correspondence-principle argument for the existence
of the cosmological solution as a physical operator was given.  Of course a
rigorous proof of the existence of this solution would require a full BRST
analysis in order to determine the actual values of the intercepts, to
verify that they were the same as those obtained from the cosmological
solution.  Such an analysis is extremely difficult and has only been
performed for the $W_3$ algebra, confirming the existence of the
cosmological solution in this case.

     In this section we shall analyse this question from another point of
view, and present very strong evidence in favour of the existence of the
cosmological solution as a physical operator.  In fact we shall argue that
the assumption that the $W_N$ string is a unitary theory implies that the
intercepts must be {\it precisely} those obtained from the cosmological
solution.

     The demonstration proceeds in two stages.  The first stage consists of
requiring that all tachyonic physical states should occur in conjugate
pairs, such that they can have non-zero norms consistent with the
momentum-conservation conditions for frozen directions discussed in subsection
5.2.  This requirement, which does not involve the assumption of unitarity,
is very natural from the point of view of the string theory.  It has the
consequence of uniquely determining the spin-3 intercept for any $W_N$
string.  For the $W_3$ string, this is therefore sufficient.  It also places
curiously-stringent constraints on the intercepts for the higher-spin currents.
The tightness of the bounds on these intercepts, which is inessential to our
further arguments, is nevertheless very intriguing.  The second stage of our
demonstration consists of demanding unitarity of the physical spectrum. This
has the effect of pinning down the values of the higher-spin intercepts
precisely, to those given by the cosmological solution.

     From the discussion in subsection 5.2, it follows immediately that if the
tachyonic physical states are to have non-zero norm, then their
momenta $(\beta_2,\ldots,\beta_N)$ must occur in conjugate pairs.  In terms
of the shifted momentum, this means that if $\vec\gamma$ is a solution of
the physical-state conditions, then $-\vec\gamma$ must be also.  In other
words if, without loss of generality, we write $\vec\gamma$ as
$$
\vec\gamma\equiv(\gamma_2,\gamma_3,\gamma_4,\ldots,\gamma_N)=
 x\Big({1\over\sqrt2},{t_3\over\sqrt6},{t_4\over\sqrt{12}},\ldots,{t_N\over
\sqrt{N(N-1)}}\Big)\acrit,\eqno(6.1)
$$
then the physical-state conditions for tachyonic states must all be {\it
even} functions of $x$.

     For the case of $W_3$, substituting (6.1) into the physical-state
conditions (3.1$b$) for tachyonic states gives
$$
\eqalign{
\omega_2&=(\acrit)^2\Big[1-\ft1{12}(t_3^2+3)x^2\Big],\cr
\omega_3&=-\acrit\,\omega_2+\ft1{108}(\acrit)^3 t_3(t_3^2-9)x^3,\cr}
\eqno(6.2)
$$
where $\acrit=7/(2\sqrt3)$.  Thus the requirement that these equations be
even in $x$ immediately implies that $\omega_3=-\acrit\,\omega_2$.  Using
the general result (3.6) for the spin-2 intercept, we therefore find
$\omega_3=-4\acrit$ for the $W_3$ string.   This agrees with the result
obtained previously from the cosmological solution (which itself agrees with
the result from the BRST analysis for $W_3$ [7], as explained earlier).  The
absence of the $x^3$ term in (6.2) requires that $t_3=0,\,\pm 3$.  It is
easy to check that these values correspond to the three conjugate pairs of
frozen momenta for the tachyons of the $W_3$ string, obtained by the action
of the Weyl group on the cosmological solution.

     Turning now to the case of the $W_N$ string, we first note that the
above discussion generalises straightforwardly to give the spin-3 intercept
of equation (3.13).  For the higher-spin currents the argument becomes more
subtle.  We shall now illustrate this with a few examples.  Let us first
consider the case of the $W_4$ string.  Substituting (6.1) into the
physical-state conditions (3.1$b$) for tachyons gives
$$
\eqalign{
\omega_2&=(\acrit)^2\Big[\ft52-\ft1{24}(2t_3^2+t_4^2+6)x^2\Big],\cr
\omega_3&=-2\acrit\,\omega_2
+\ft1{216}(\acrit)^3(2t_3+t_4)(t_3-t_4+3)(t_3-t_4-3)x^3,\cr
\omega_4&=-\ft9{16}(\acrit)^4 -\ft34(\acrit)^2\omega_2 -\ft32\acrit\,\omega_3
+\ft1{6912}(\acrit)^4 t_4(4t_3-t_4)(t_4+2t_3-6)(t_4+2t_3+6)x^4,\cr}
\eqno(6.3)
$$
where $\acrit=9/(2\sqrt5)$.  From the single condition needed to make these
equations of even order in $x$, it follows that there are three 1-parameter
families of solutions at this stage.  Using the value for the spin-2
intercept given by (3.6) we find that $\omega_3$ is then uniquely determined,
as  mentioned above, to be $-20\acrit$, but that $\omega_4$ is a function of
the free parameter, say $t_3$, for each of the three families. The range of
$\omega_4$ is the same for each family, and spans a remarkably small
interval:
$$
81.89859375={524151\over6400}\le\omega_4\le{524176\over6400}= 81.9025.
\eqno(6.4)
$$
The cosmological solution corresponds to $\omega_4=819/10$.  We have no
explanation, beyond the superficial one, for why the allowed range in (6.4)
is so small.  It seems however that this is a generic feature of $W_N$
strings, which we have also observed for $W_5$ and $W_6$, suggesting a deeper
structure that has  still to be elucidated.

     We have just seen that requiring that the tachyonic physical states of
the $W_4$ string arise in conjugate pairs determines the spin-3 intercept
uniquely, and restricts the spin-4 intercept to lie in the interval given in
(6.4).  We shall now consider the consequences of unitarity.  We discussed
this in subsection 5.2 for the case where the intercepts were assumed to take
the values determined by the cosmological solution.  We saw in particular
that, under this assumption, the physical states of the $W_N$-string theory
of the form (5.6) have a set of $L_0^{\rm eff}$ values given in  (5.7) that
lie in the interval $3(N+2)/(4(N+1))\le L_0^{\rm eff}\le 1$ of equation
(5.9),  including states that saturate the lower and the upper bounds.  As
we shall  now show, if the $\omega_4$ intercept for $W_4$ is not given by
its  cosmological value, then some states will violate these unitarity
bounds.

      The values of $L_0^{\rm eff}$ for these states are given by
substituting their (frozen) $\beta_2$ values into (5.7) (there are no extra
$X^\mu$ coordinates in our present discussion).  In terms of the shifted
momentum given by (6.1), we therefore have
$$
L_0^{\rm eff}=\ft{81}{80}(1-x^2)\eqno(6.5)
$$
for the $W_4$ case.  Since we already know the value of $\omega_2$ from (3.6),
the first equation in (6.3) may be used to express the  free parameter $t_3$ in
terms of $x^2$, and hence, using (6.5), in terms of  $L_0^{\rm eff}$.  Thus for
each family we may express the spin-4 intercept  $\omega_4$ as a function of
$L_0^{\rm eff}$.  Two of the three families give  the same result,
$$
\omega_4(L_0^{\rm eff})=\omega_4^{\rm cosmo} -4\big(1-L_0^{\rm eff}\big)
\big(L_0^{\rm eff}-\ft{77}{80}\big),\eqno(6.6)
$$
and the third family gives
$$
\omega_4(L_0^{\rm eff})=\omega_4^{\rm cosmo} +\big(1-L_0^{\rm eff}\big)
\big(L_0^{\rm eff}-\ft{9}{10}\big).\eqno(6.7)
$$
In these equations $\omega_4^{\rm cosmo}$ denotes the cosmological value,
$819/10$, for the spin-4 intercept for $W_4$.  Equation (6.6) shows that
there is a state that has $L_0^{\rm eff}=1$ when $\omega_4=\omega_4^{\rm
cosmo}$ and has $L_0^{\rm eff}>1$ when $\omega_4$ is {\it larger} than
$\omega_4^{\rm cosmo}$.  On the other hand, equation (6.7) shows that there
is another state that has $L_0^{\rm eff}=1$ when $\omega_4=\omega_4^{\rm
cosmo}$ and has $L_0^{\rm eff}>1$ when $\omega_4$ is {\it smaller} than
$\omega_4^{\rm cosmo}$.  What is happening is that the degeneracy of the
set of physical states at a given $L_0^{\rm eff}$ value described in subsection
4.4 is (partially) lifted when $\omega_4$ is not equal to its cosmological
value $\omega_4^{\rm cosmo}$.  For some such states, $L_0^{\rm eff}$
increases with {\it increasing} $\omega_4$, while for other states $L_0^{\rm
eff}$ increases with {\it decreasing} $\omega_4$.  Therefore by considering
the full set of originally-degenerate $L_0^{\rm eff}=1$ states, we see that
unitarity will be violated for any value of $\omega_4$ except the
cosmological value.\footnote{$^\dagger$}{\tenfoot In fact the requirement of
unitarity by itself is sufficient to determine {\it all} the intercepts,
including the spin-2 intercept which we calculated in section 3 by a
different method.  One simply uses the same procedure that we have described
for $\omega_4$ to express $\omega_2$ as a function of $L_0^{\rm eff}$, and
sees that the unitarity bounds are violated unless $\omega_2$ is given by
(3.6).}

     We have also checked completely that the same conclusions hold in the case
of the $W_5$ string. Since the general pattern is very similar to the $W_4$
case, we shall be very brief about it. The intercepts of the spin-4 and spin-5
currents again depend on one free parameter and their ranges are, once more,
remarkably small:
$$
\eqalign{
499.59555\cdots={1798544\over3600}&\le\omega_4\le{1798569\over3600}=499.6025 ,
\cr
-515.205=-{927369\over1800}&\le{\omega_5\over\acrit}\le-{927344\over1800}=
-515.19111\cdots\, .\cr }
\eqno(6.8)
$$
The unitarity bounds (5.9) for $N=5$ are violated except when the intercepts
take the values determined by the cosmological solution.

     If $N>5$, then the situation is more complicated.  In the case of $W_6$,
for example, the higher-spin intercepts now depend on two free parameters and
lie again in extremely small ranges. Although we have not analysed the general
case in detail, we expect that only the cosmological solution will give a
unitary theory.

\bigskip\bigskip
\noindent {\bf 7. Conclusions}
\bigskip

     In this paper we have studied the physical spectrum of $W_N$-string
theories.  Starting from the Miura transformation for $su(N)$, we derived an
explicit formula giving the currents of $W_N$ in terms of those of
$W_{N-1}$, together with one extra free scalar.  Applying this recursively
leads to realisations of $W_N$ in terms of $(N-2)$ free scalar fields
$(\varphi_3,\ldots,\varphi_N)$ and an arbitrary energy-momentum tensor.  By
taking this energy-momentum tensor to be realised in terms of additional
free scalar fields, one has the starting point for a $W_N$-string theory.
In order to study the physical-state conditions for this $W_N$ string, we
used a method based on a unitarity argument to determine the intercepts
of the $W_N$ currents, avoiding the necessity of performing the complete
BRST analysis of the $W_N$ gauge theory.  This allowed us to derive explicit
formulae for these intercepts.

     Using these values of the intercepts, we gave a construction of all the
tachyonic physical states for the general $W_N$ string.  All these
states can be obtained by acting on a particular physical state, the
cosmological solution, with the Weyl group of $su(N)$. We also constructed all
the level-1 and level-2 physical states in some specific examples. Our results
indicate that the physical spectrum of the $W_N$ string bears a strong
resemblance to ordinary Virasoro string theory, but with a non-critical value
of the central charge (1.2) and a discrete set of intercepts, given by (4.14)
and (4.18),  which includes 1, together with other values.  These values for
the
central charge and the intercepts are very suggestive of a connection between
$W_N$ strings and minimal models.  The physical states of the $W_N$ string
can, in a certain sense, be viewed as ``gravitational dressings'' of the
primary fields of the corresponding minimal model [2].  The precise nature
of this connection is, however, still a mystery.

     Our finding that unitarity requires the values of the intercepts for
the $W_N$ currents to be those given by the cosmological solution is one of
the crucial results of this paper.  We studied the no-ghost theorem by
analysing level-1 and level-2 states. These, and indeed as we have seen {\it
all} higher-level states, can be divided into two categories, namely those
that comprise excitations {\it only} in the (unfrozen) $(\varphi_2,X^\mu)$
directions, and the remaining ones which include excitations in the (frozen)
$(\varphi_3,\ldots, \varphi_N)$ directions.  The first category describes
the higher-level states of the effective Virasoro-string theory alluded to
above; it is the unitarity of these states that determines the values of the
$W_N$ intercepts. The second category, as we showed in various examples,
describes higher-level physical states that {\it all} have zero norm, since
they do not occur in conjugate pairs, implying that the two-point functions
of any two of these states vanishes by momentum conservation.  Since only
the first category of higher-level states contributes to the physical
spectrum of the $W_N$ string, the derivation of the  {\it entire} physical
spectrum of the $W_N$ string reduces to finding the physical spectra of a
set of Virasoro-type string theories,  with the non-standard values of the
central charge and intercepts given above.  Thus we have found the complete
spectrum of the $W_N$ string.

     We have shown that $W_N$-string theory reduces to effective
Virasoro-string theories with $(D+1)$ coordinates $X^\mu$ and $\varphi_2$. For
$N\ge 3$, the coordinate $\varphi_2$ has a non-vanishing background charge
$Q$, given in (4.10).   Owing to this background charge, the theory does not
have $(D+1)$-dimensional target-space Poincar\'e invariance. Amongst other
things, this makes the definition of a $(D+1)$-dimensional mass ambiguous.
This problem resolves itself by taking $D$ to be greater than 24, since then
the background charge becomes imaginary. As discussed in [3] for the $W_3$
string, this implies that the $\varphi_2$ coordinate has to live on a
circle, since the functional integral becomes periodic in $\varphi_2$ with
period $\pi/|Q|$. Therefore its momentum component $\beta_2$ is quantised,
and is given by
$$
\beta_2 =2 m i |Q|, \qquad m \in \Z. \eqno(7.1)
$$
(Recall that in our conventions, the momenta are imaginary.)
Because $\varphi_2$ is then compactified, the remaining coordinates $X^\mu$
describe a $D$-dimensional Minkowski spacetime {\it \`a la} Ka{\l}uza-Klein.
Since this spacetime {\it is} Poincar\'e invariant, the definition of
$D$-dimensional mass is now unambiguous and given by ${\cal M}^2=\beta_\mu
\beta^\mu$. Using (5.7), (4.14), (4.10), (4.18) and (7.1) this can be
rewritten for the $W_N$ string as
$$
{\cal M}^2= -2 +2n + {k^2-1\over 2N(N+1)} + \ft13 m(m-1)\Big[D-24
-{6\over N(N+1)}\Big], \eqno(7.2)
$$
where $n$ is the level number of the physical state, $m$ is the Ka{\l}uza-Klein
mode number and $k$ is an integer labelling the diagonal entries of the Kac
table of the relevant minimal model, satisfying $1\le k \le N-1$. In particular
we see from this that the $k=1$ case with $m=0$ or $m=1$ has precisely the
mass spectrum of ordinary Virasoro-string theory and includes, therefore, a
massless vector at level $n=1$.  Curiously, some
sporadic cases develop extra massless physical states, occurring at level
$n=0$. These massless tachyons arise at $D=25$, $m=-2,3$, $k=5$ for $W_N$
strings with $N\ge 6$,  and at $D=27$, $m=-1,2$, $k=3$ for  $N\ge 4$.

     The issue of the existence of massless states in the spectrum of the
$W_N$  string had already been raised some time ago.  Since the spin-2
intercept (3.6) for the $W_N$ string increases with $N$, naive expectations
might suggest that the $({\rm mass})^2$ of physical states would
correspondingly be decreased, opening the possibility of having massless
higher-spin states.  In fact, as seen in [3] and this paper, the effect of
having momentum components frozen by the $W_N$ constraints is that the
relevant quantity that determines the masses of the physical states is
$L_0^{\rm eff}$ rather than the spin-2 intercept (3.6).  Since $L_0^{\rm
eff}\le 1$, it follows that the values of $({\rm mass})^2$ are either
the same as those for the usual Virasoro string (when $L_0^{\rm eff}=1$) or
shifted {\it upwards}.  The original version of [3] mistakenly discarded the
physical states of the $W_3$ string with $L_0^{\rm eff}=1$, by imposing an
over-stringent requirement of hermiticity of the $W_3$ currents.  This led to
the erroneous conclusion that the $W_3$ string did not contain massless
states at all. (Further discussion of this issue may be found in [10], and
the revised version of [3].)  In fact, as we have just seen in the previous
paragraph, $W_N$ strings do contain massless states.  In particular, the
massless level-1 states will give rise to a massless graviton in the case of
a closed $W_N$ string. However, neither the open nor the closed $W_N$ string
has massless states with spins higher than 2.

     The next step in the understanding of the $W_N$ string is to construct an
interacting theory.

\bigskip\bigskip
\centerline{\bf ACKNOWLEDGMENTS}
\bigskip

     We are grateful to Kelly Stelle for discussions on the r\^ole of the
momentum-conservation law in frozen directions for norm calculations.
Related work will appear in [10].  We have made extensive use of the
Mathematica package {\sl OPEdefs} version 2.1 for calculating
operator-product expansions [11]. Stany Schrans is indebted to the Center
for Theoretical Physics, Texas A\&M University, for hospitality, and to the
Belgian National Fund for Scientific Research for a travel grant.

\bigskip\bigskip
\centerline{\bf NOTE ADDED}
\bigskip

     After this paper was completed, we encountered a paper that has some
overlap with our work [12].  It discusses the relation between the
tachyonic spectrum of the $W_N$ string and the diagonal states of the
corresponding minimal model, although the r\^ole of the Weyl group as the
organising symmetry is not found.

\np

\singlespace
\centerline{\bf REFERENCES}
\frenchspacing
\bigskip

\item{[1]}C.N.\ Pope, L.J.\ Romans and K.S.\ Stelle, {\sl Phys.\
Lett.}\ {\bf 268B} (1991) 167;  {\sl Phys.\ Lett.}\ {\bf 269B} (1991) 287.

\item{[2]}S.R.\ Das, A.\ Dhar and S.K.\ Rama, {\sl Mod.\ Phys.\ Lett.}\
{\bf A6} (1991) 3055; ``Physical states and scaling properties of $W$
gravities and $W$ strings,'' preprint, TIFR/TH/91-20.

\item{[3]}C.N.\ Pope, L.J.\ Romans, E.\ Sezgin and K.S.\ Stelle, ``The $W_3$
String Spectrum,'' preprint CTP TAMU-68/91, to appear in {\sl Phys.\ Lett.\ }
{\bf B}.

\item{[4]}V.A.\ Fateev and S.\ Lukyanov,  {\sl Int.\ J.\ Mod.\  Phys.}\ {\bf
A3} (1988) 507.

\item{[5]}L.J.\  Romans, {\sl Nucl.\  Phys.}\ {\bf B352} (1991) 829.

\item{[6]}A.\ Deckmyn and S.\ Schrans, ``$W_3$ constructions on affine Lie
algebras,'' Preprint-KUL-TF-91/33, to appear in {\sl Phys.\ Lett.\ }
{\bf B}.

\item{[7]}J.\ Thierry-Mieg, {\sl Phys.\ Lett.}\  {\bf 197B} (1987) 368.

\item{[8]}H.\ Lu, C.N.\ Pope, X.J.\ Wang and K.W.\ Xu, ``Anomaly Freedom and
Realisations for Super-$W_3$ Strings,''  preprint CTP TAMU-85/91.

\item{[9]}V.A.\ Fateev and A.\ Zamolodchikov, {\sl Nucl.\  Phys.}\  {\bf
B280} (1987) 644.

\item{[10]}H.\ Lu, C.N.\ Pope and K.S.\ Stelle, in preparation.

\item{[11]}K. Thielemans, {\sl Int.\ J.\ Mod.\ Phys.}\  {\bf C2} (1991) 787.

\item{[12]}S.K.\ Rama, {\sl Mod.\ Phys.\ Lett.}\ {\bf A6} (1991) 3531.

\bye